\documentclass{article}

\usepackage{PRIMEarxiv}

\usepackage[utf8]{inputenc} 
\usepackage[T1]{fontenc}    
\usepackage{hyperref}       
\usepackage{url}            
\usepackage{booktabs}       
\usepackage{amsfonts}       
\usepackage{nicefrac}       
\usepackage{microtype}      
\usepackage{lipsum}
\usepackage{fancyhdr}       
\usepackage{graphicx}       
\graphicspath{{media/}}     

\usepackage{amsmath}

\usepackage{algpseudocode}
\usepackage{algorithm}

\usepackage[Gray,squaren,thinqspace,thinspace]{SIunits}
\usepackage{tikz}
\usepgflibrary{shapes.multipart}
\usetikzlibrary{arrows,backgrounds,arrows.meta}

\title{Optimizing the Traversal Time for Gantry Trajectories for Proton Arc Therapy Treatment Plans
}

\author{
  Viktor Wase, Otte Marthin, Albin Fredriksson, Anton Finnson \\
  RaySearch Laboratories AB\\
  Stockholm \\
  Sweden\\
  \texttt{\{firstname.lastname\}@raysearchlabs.com} \\
}

\begin{document}
\maketitle

\begin{abstract}
\noindent {\bf Background:} Proton arc therapy is an emerging radiation therapy technique where either the gantry or the patient continuously rotates during the irradiation treatment. One of the perceived advantages of proton arc therapy is the reduced treatment time, but it is still unclear exactly how long these treatment times will be, given that no machine capable of its delivery is available at the market at the time of writing.\\
{\bf Purpose:} We introduce the algorithm Arc Trajectory Optimization Method (ATOM), which aims to determine an efficient velocity profile for the gantry for rapid delivery of a given proton arc treatment plan.\\
{\bf Methods:} ATOM computes the trajectory with the shortest delivery time while ensuring there is enough time to deliver all spots in each energy layer and switch energy between layers. The feasibility of the dynamic gantry movement was assured by enforcing maximum and minimum limits for velocity, acceleration, and jerk. This was achieved by discretizing the gantry velocity and combining the A* algorithm with the open-source motion generation library \textit{Ruckig}. The algorithm was tested on a synthetic data set as well as a liver case, a prostate case and a head and neck case.\\
{\bf Results:} Arc trajectories for plans with 360 energy layers were calculated in under a second using 256 discrete velocities.\\
{\bf Conclusions:} ATOM is an open-source C++ library with a Python interface that rapidly generates velocity profiles, making it a highly efficient tool for determining proton arc delivery times, which could be integrated into the treatment planning process.\\

\end{abstract}

\keywords{ Radiotherapy \and Ion Therapy \and Proton Arc}






\pagenumbering{arabic}
\setcounter{page}{1}
\pagestyle{fancy}

\section{Introduction}
Proton arc therapy (PAT) is an emerging radiation therapy treatment technique delivering the protons in pencil beam scanning mode while the gantry moves, as opposed to intensity modulated proton therapy (IMPT) where the gantry is stationary while irradiating. Optimization algorithms for creating PAT treatment plans often try to reduce the delivery time indirectly by reducing the number of energy switches to higher energies \cite{elsa, sparc, spat}. The number of upward energy switches is considered a good predictor of beam delivery time (BDT), since an upward switch tends to be slower than a downward energy switch for most delivery systems. Zhao et al. \cite{spot_opt} modeled BDT as proportional to the number of spots that also increases considerably in PAT plans and derived a method to enforce sparsity in the spot selection. Another group developed a technique to generate a Pareto front balancing delivery time and plan quality for proton arcs, using an approximate static BDT model taking into account energy switches and layer irradiation time \cite{pareto}. A common dynamic BDT model assumes a constant gantry rotation speed of $6\degree\per\second$, plus the time added by irradiation segments and energy switching \cite{sparc, lung, hippo}.

In this paper we introduce an algorithm that can be used to design the gantry velocity curve of a proton arc plan such that the delivery time is minimized. The algorithm can be used in order to decide the velocity profile of a plan after the plan has been designed. Alternatively, it can be used repeatedly during the proton arc plan optimization process to take delivery time into account when creating the plan. The algorithm is tested on data meant to simulate a PAT plan, but the algorithm works for other types of ion arcs as well, since the overall delivery framework is very similar \cite{sharc}. The algorithm considers limits on maximum gantry velocity, acceleration as well as jerk (the derivative of acceleration). Including the jerk is important to create a smooth velocity profile while limiting the machine wear \cite{obravibrationer}.

At the time of writing there is, to our knowledge, no published algorithm that designs the velocity curve of the gantry for an ion arc treatment plan. The inertia could have a significant impact on BDT since the gantries are incredibly heavy and the delivery time per energy layer varies a lot. Changes in inertia is important to consider in the cases when the patient is moved and the gantry is static, since perturbations to the position of the patient should be minimized. An algorithm that is fast and general enough to be used by multiple research teams and multiple machines would greatly improve the effort required to compare the reported delivery times in different research papers. The present paper seeks to address this need by proposing an algorithm that meets these criteria.
\section{Methods and Material}
\subsection{Proton Arc Preliminaries}
\begin{figure}[H]
    \centering
    \includegraphics[scale=0.35]{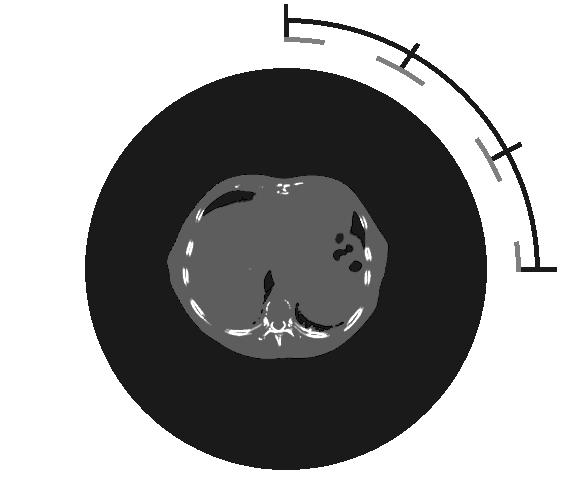}
    \caption{Example of a proton arc plan, covering a quarter of a full revolution and consisting of four energy layers. The gray lines represent the gantry angle windows.}
    \label{fig:expl}
\end{figure}

The main idea of proton arc is to move the gantry while irradiating. It is to IMPT what VMAT is to IMRT. A proton arc plan is defined in much the same way as a regular IMPT plan, but with significantly more beams. However, much like in VMAT planning, these are not refered to as \textit{beams}, but rather \textit{gantry angles} or simply \textit{directions}. Each such gantry angle corresponds to a \textit{gantry angle window}. The gantry angle window is a small angular range of the arc, and each such window has a set of spots assigned to it, all of which have to be delivered while the gantry angular position is located inside the gantry angle window. The idea is that this window should be small enough that it does not matter at which exact angle a spot is delivered, as long as it is within the specified window. The dose perturbation should be small enough to be negligible. See Figure \ref{fig:expl} for an illustration of this concept.

Throughout this paper we assume that each specified gantry angle contains exactly one energy layer. However, this assumption is made without loss of generality to simplify the terminology. If several energy layers are to be delivered within one gantry window, one can simply consider the combination of these energy layers and treat them as a single energy layer with irradiation time equal to the sum of the total irradiation time of all layers and the internal energy layer switching times.
\subsection{Model}
In this paper we introduce the Arc Trajectory Optimization Method (ATOM), which is capable of producing optimized velocity profiles for ion arc plans, given the gantry angles and irradiation times of all energy layers, as well as the energy switching times between them. The goal of the algorithm is to create a velocity profile such that the total delivery time is minimized, while fulfilling all conditions regarding how much time the gantry needs to spend in and between the gantry angle windows in order to properly deliver the plan.



In order to keep the deviation between the planned and delivered dose small, we center the irradiation of each gantry angle window by keeping the gantry velocity fixed while irradiating and starting/stopping the irradiation equally long before/after the center of the gantry angle window. This means that the acceleration is zero during irradiation. Allowing non-zero acceleration during irradiation could possibly allow for faster delivery, but may increase the difference between planned and delivered plan. For example, the irradiation might become systematically skewed towards the latter part of the gantry angle windows, effectively rotating the dose distribution. Moreover, it increases the dimensionality of the problem, since both acceleration and velocity must be determined for each gantry angle window.

The problem to be solved is thus that of choosing a sequence of velocities at which to start irradiation. The allowed velocities for a given gantry angle are dependent on the previous and following velocities, and are also restricted by the limitations on acceleration and jerk. This gives the problem a combinatorial structure well suited for a dynamic programming formulation. Such a formulation does, however, require a discretization of the set of possible velocities for the gantry at start of irradiation. By repeating experiments with different discretization resolutions we choose a suitable precision with which to specify these velocities, so that the gain from further precision is negligible compared to the increase in computation time.

A final assumption in the ATOM solution is to not allow the gantry to switch direction of travel. This enables quicker solution of the dynamic programming problem, and while this might increase the optimal delivery time slightly, we find it reasonable to believe that clinics in practice would want to enforce this rule, as a heavy proton gantry that switches direction back and forth throughout the same arc seems problematic with regards to safety and QA. Note that we investigate the effects of these assumptions on both delivery time and the algorithm's computational time in our experiments.

\subsection{Mathematical Formulation}
The problem of optimizing total delivery time can be subdivided into two main problems. The first problem consists of finding the minimal travel time between any two arc states, consisting of position, velocity, acceleration, and jerk. Having solved this first problem, the second problem becomes that of constructing the fastest path through a sequence of arc states, utilizing that we know the fastest solution between any two states. 

Denote the angle span between the edges of two adjacent gantry angle windows by $\Delta x$. Given an initial velocity $a$ and a final velocity $b$, let $t_i(a, b)$ denote the traversal time of the fastest velocity profile over the angle span $\Delta x$ that satisfies the machine constraints and the required time for energy switching between the $i$th and $(i+1)$th energy layer, as well as the initial and final velocities. We will also need to consider the irradiation time of the energy layers. Define $t^{\textrm{irr}}_i$ as the time required to deliver all spots in the $i$th energy layer.

\begin{figure}
\begin{center}
    \begin{tikzpicture}
    	\tikzstyle{place}=[circle, draw=black, minimum size = 4mm]
  		\draw node at (-3, -2.25*1.8) [place, minimum size=1.7cm] (start) {$n_{1,1}$};
            
    	\foreach \x in {1,...,2}
    		\draw node at (0, -\x*1.8) [place, minimum size=1.7cm] (first_\x) {$n_{2,\x}$};
    	\foreach \x in {1,...,3}
    		\fill (0, -4.55 -\x*0.3) circle (1pt);
    	\draw node at (0, -4*1.8 + 0.45) [place, minimum size=1.7cm] (first_n) {$n_{2,M}$};
    	
    	\foreach \x in {1,...,2}
    		\node at (3, -\x*1.8) [place, minimum size=1.7cm] (second_\x){$n_{i,\x}$};
    	\foreach \x in {1,...,3}
    		\fill (3, -4.55 -\x*0.3) circle (1pt);
    	\draw node at (3, -4*1.8 + 0.45) [place, minimum size=1.7cm] (second_m) {$n_{i,M}$};
    	
            \foreach \x in {1,...,2}
    		\node at (6, -\x*1.8) [place, minimum size=1.7cm] (fourth_\x){$n_{N-1,\x}$};
            \foreach \x in {1,...,3}
    		\fill (6, -4.55 -\x*0.3) circle (1pt);
    	\node at (6, -4*1.8 + 0.45) [place, minimum size=1.7cm] (fourth_m) {$n_{N-1,M}$};

        \draw node at (9, -2.25*1.8) [place, minimum size=1.7cm] (end) {$n_{N,1}$};

    	\foreach \i in {1,...,2}
  			\draw [-{Latex[length=2mm, width=2mm]}] (start) to (first_\i);
		\draw [-{Latex[length=2mm, width=2mm]}] (start) to (first_n);

    	\foreach \i in {1,...,2}
    		\foreach \j in {1,...,2}
    			\draw [-{Latex[length=2mm, width=2mm]}] (first_\i) to (second_\j);
    	\foreach \i in {1,...,2}
    		\draw [-{Latex[length=2mm, width=2mm]}] (first_\i) to (second_m);
    	\foreach \i in {1,...,2}
    		\draw [-{Latex[length=2mm, width=2mm]}] (first_n) to (second_\i);
	\draw [-{Latex[length=2mm, width=2mm]}] (first_n) to (second_m);
    	
  	\foreach \i in {1,...,2}
		\foreach \j in {1,...,2}
    			\draw [-{Latex[length=2mm, width=2mm]}] (second_\i) to (fourth_\j);
	\foreach \i in {1,...,2}
    		\draw [-{Latex[length=2mm, width=2mm]}] (second_\i) to (fourth_m);
	\foreach \i in {1,...,2}
    		\draw [-{Latex[length=2mm, width=2mm]}] (second_m) to (fourth_\i);
    	\draw [-{Latex[length=2mm, width=2mm]}] (second_m) to (fourth_m);

    	\foreach \i in {1,...,2}
  			\draw [-{Latex[length=2mm, width=2mm]}] (fourth_\i) to (end);
		\draw [-{Latex[length=2mm, width=2mm]}] (fourth_m) to (end);
    	
    	\node at (-3, -8) [black, ] {First angle};
    	\node at (0, -8) [black, ] {Second angle};
    	\node at (3, -8) [black, ] {Angle $i$};
    	\node at (6, -8) [black, ] {Angle $N-1$};
    	\node at (9, -8) [black, ] {Last angle};
    \end{tikzpicture}
    \caption{Illustration of a network of for determining velocity profiles. Each column represents a gantry angle that corresponds to the central angle of an energy layer, and each row $j$ represents a discrete velocity $s_j$. The cost between two nodes $n_{i,j}$ and $n_{i+1, k}$ is equal to $t(s_j, s_k) + t^{\textrm{irr}}_i$.}
    \label{fig:dynp}
\end{center}
\end{figure}
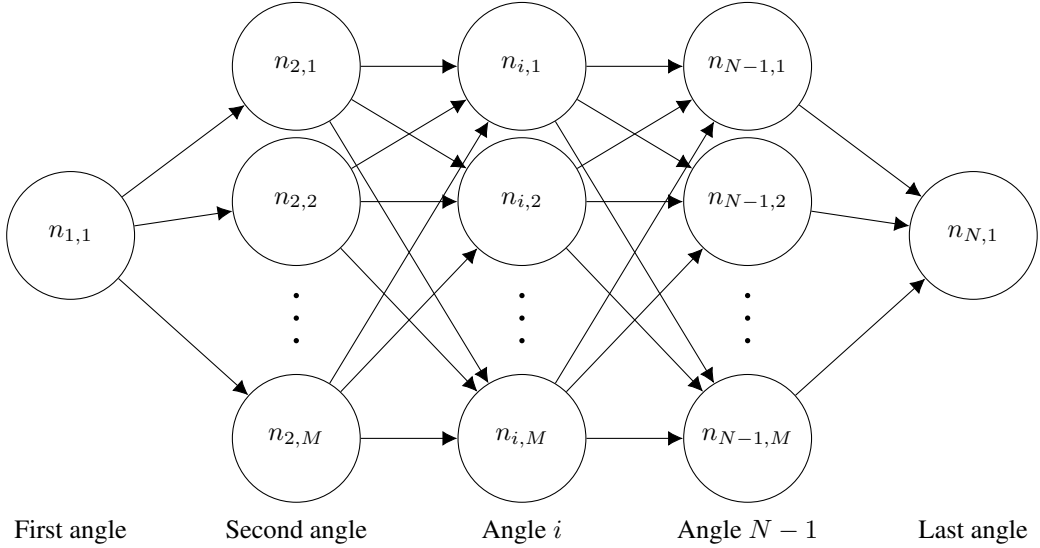

If there are $N$ gantry angle windows, the velocity profile problem is that of finding the velocities $v_1, \ldots, v_N$ of all gantry angle windows such that the total trajectory time 
\[
\sum_{i=2}^N t_i(v_{i-1}, v_i) + \sum_{i=1}^N t^{\textrm{irr}}_i
\]
is minimized. To this end, we construct a directed graph of nodes, each corresponding to an gantry-angle-window-and-velocity state. Let there be $M$ discrete velocities $s_1,\ldots, s_M$ to choose from, and each node is denoted by $n_{i,j}$, where the node for $i=2,\ldots,N-1$ and $j=1,\ldots,M$ corresponds to the $i$th gantry angle window and the $j$th discrete velocity ($s_j$), and the starting node $n_{1,1}$ and the last node $n_{N,1}$ correspond to a velocity of zero at the first and last gantry angle window, respectively, i.e., $v_1 = v_N = 0$. The nodes have directed edges with connections corresponding to moving the gantry one angle-step forward, to the next energy layer. Each edge has an associated cost; the cost between node $n_{i,j}$ and $n_{i+1,k}$ is given by $t_i(s_j, s_k) + t^{\textrm{irr}}_i$, the transition time between the corresponding two states plus the irradiation time for the gantry angle window we are leaving. Note that the function $t_i(v_{i-1}, v_i)$ implicitly depends on $\Delta x$, $t^{irr}_{i-1}$ and $t^{irr}_{i}$, but they are omitted for clarity since they are constants and treated as implicit to the functions $t_i$. Further note that the irradiation time of the last layer is ignored in the problem, and simply added later as a constant. An illustration of this graph is shown in Figure~\ref{fig:dynp}.

\begin{figure}
    \centering
    \includegraphics[scale=0.8]{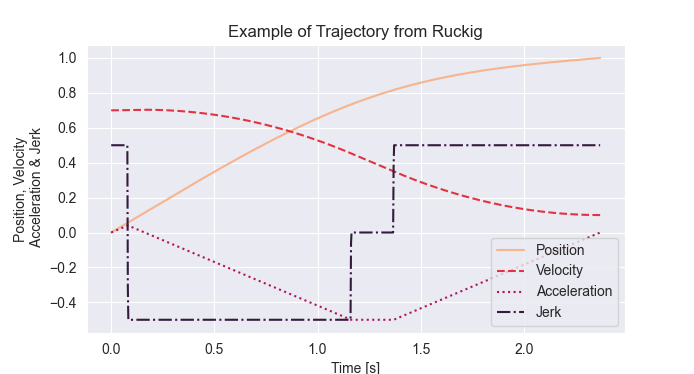}
    \caption{Example of a trajectory generated by \textit{Ruckig}. The initial velocity is $0.7 \degree\per\second$ and the final is $0.1 \degree\per\second$, with a delta angle of $1\degree$ and a maximum/minimum jerk of $ 0.5 \degree\per\second^3$, maximum/minimum acceleration of $0.5 \degree\per\second^2$, and maximum velocity of $5 \degree\per\second$.}
    \label{fig:traj}
    \vspace{0.7cm}
\end{figure}

Determining the cost of each edge in the graph amounts to solving a sub-problems to find the optimal trajectory between two velocities, while covering a given distance. This is a well studied problem~\cite{traj_type, ruckig, type_iv}. Berscheid et al. \cite{ruckig} published an open-source C++ library called \textit{Ruckig} that quickly solves this problem to optimality under constraints on maximum and minimum velocity, acceleration and jerk. In fact, this library can solve the more general problem of having non-zero boundary accelerations, but is still significantly faster than other algorithms specifically written to find trajectories with zero boundary acceleration~\cite{ruckig}. Therefore, we use \textit{Ruckig} to calculate the duration $t_i(s_j, s_k)$ of the trajectories between the gantry angle windows. The library also supports enforcing a minimum travel time, which means that we can make sure that the trajectory is always slower than the required time for energy layer switching. An example trajectory is illustrated in Figure \ref{fig:traj}.

The way \textit{Ruckig} is adapted to the problem of computing the optimal transition time between the nodes in consecutive layers of the graph is described in detail in Algorithm \ref{algo:1}. Note that the cost is set to infinity if no feasible trajectory exists between two nodes. With this construction, the lowest-cost path through the graph corresponds to the velocity sequence of the time-optimal arc.

\begin{algorithm}[H]
\caption{Transition Time}
    \begin{algorithmic}
    \State{\textbf{Inputs:}}
    \State{$v_0, v_1$ \Comment{The initial and final velocities.}}
    \State{$t_0, t_1$ \Comment{The irradiation times of the previous and next energy layers.}}
    \State{$t_s$ \Comment{The time required to switch from the previous energy layer to the next one.}}
    \State{$\Delta x$ \Comment{The angular distance between the middle of the previous and next gantry angle.}}
    \State{\textbf{Procedure:}}
    \State{window$_0$ := $v_0 t_0$ \Comment{Shrink the gantry angle windows to the smallest possible size.}}
    \State{window$_1$ := $v_1 t_1$}

    \If{window$_0 > $ max gantry angle window size OR window$_1 > $ max gantry angle window size}
    \State{\Return $\infty$}
    \EndIf

    \State{angleSpan := $\Delta x - \frac{1}{2}($window$_0 + $window$_1)$ }
    \State{trajectory := calculateTrajectory($v_0, v_1,$ angleSpan, $t_s$) \Comment{Calculated using \textit{Ruckig}.}}
    \If{trajectory.isInfeasible}
    \State{\Return{ $\infty$}}
    \EndIf
    \State{\Return{trajectory.transitionTime + $t_0$}}

    \end{algorithmic}
\label{algo:1}
\end{algorithm}

Finding the shortest path in a directed graph can be done using a variety of algorithms. Here, the trajectory problem is solved to optimality using the A* search algorithm \cite{astar}.

The A* algorithm requires a heuristic to estimate the remaining path cost from each node to the end node. If this heuristic estimation is smaller than or equal to the true value for all nodes, then the A* is guaranteed to find the optimal solution \cite{astar}. We use the sum of the remaining energy layer switching times plus the sum of the remaining irradiation times as our heuristic. This means that all nodes that share gantry angle, but not necessarily gantry velocity, will share heuristic value. This value is always smaller than or equal to the remaining delivery time, which means that the found path will be optimal.

In order to reduce the running time of the algorithm we prune the graph. Given a current velocity $v_i$ and delta angle $\Delta x$ we calculate the maximum and minimum velocity that can possibly be reached for the next gantry angle, assuming that there is no limit on jerk (i.e. acceleration can change instantaneously) and that the upcoming delivery window is of width 0. From these assumptions and using the kinematics equations of motion, we get
\begin{align*}
   v_\text{low} := \sqrt{ v_i^2-2  a_{\max} \Delta x},\quad v_\text{high} := \sqrt{v_i^2+2  a_{\max} \Delta x}.
\end{align*}
When searching through velocity transitions $v_i \to v_{i+1}$ we then only consider $v_{i+1}$ within $[v_\text{low},v_\text{high}]$. If $v_\text{low}$ is imaginary, then we simply use 0 as a lower bound.

In summary, ATOM constructs a graph representing all gantry angle windows and a discretized set of possible velocities during the irraditation; then computes the traversal time between each pair of gantry angle windows and each pair of discrete velocities using Algorithm~\ref{algo:1}; and finally constructs a sequence of gantry velocities by solving for the shortest path through the graph.


The code is publicly available for download from the GitHub repository https://github.com/raysearchlabs/ArcTrajectoryOptimizationMethod under the liberal MIT license, and can be used by anyone for research purposes. It is available in both C++ and Python. The run times reported in this paper are based on the C++ version, running on an Intel Core i9-10940X CPU and using Windows 10. The code was not parallelized.

\subsection{Experiments}
\subsubsection{Synthetic Data}
The algorithm was run on a synthetic dataset consisting of 1000 plans. Each plan had 360 energy layers, with one layer every $1\degree$ along the arc, each with a gantry angle window of width $1\degree$. The energy layer switching times were generated independently with a 90\% chance of taking 0.5 s and a 10\% chance of taking 5 s, representing down-switches and up-switches. The irradiation time per layer is uniformly distributed between 0 and 1.26 s in order to get the irradiation to represent about 40\% of the static time.

The algorithm was tested with two different sets of machine parameters: one with a small range of allowed accelerations (acceleration-limited machine parameters) and a second one allowing for more acceleration while limiting the jerk more (jerk-limited machine parameters). For each set of machine parameters the algorithms was evaluated both with and without using the pruning strategy described above. See Table \ref{tab:my_label1}.

\begin{table}[H]
    \centering
    \begin{tabular}{c|c|c|c}
             & $v_{\text{max}}$ & $a_{\text{max}}$ & $j_{\text{max}}$ \\
            \hline
         Jerk Limited Machine Parameters & 5.0 & 0.5& 0.5 \\
        Acceleration Limited Machine Parameters & 5.0 & 0.25& 1.0

    \end{tabular}
    \caption{The algorithm was tested on two datasets, based on these machine parameters.}
    \label{tab:my_label1}
\end{table}

The tests were repeated for varying velocity discretization, as well, but with 100 random plans instead of 1000 to make it faster. This was done to get a sense of how to balance the running time and the accuracy of the algorithm.

\subsubsection{Patient Data}
\begin{figure}
    \centering
    \includegraphics[scale=1]{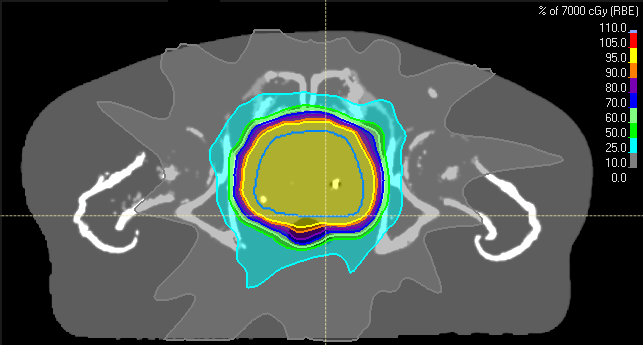}
    \caption{Proton Arc plan for the prostate case, with 180 energy layers and an arc spanning the full revolution.}
    \label{fig:prostate}
\end{figure}

The algorithm is also tested on three patient cases. The data is a liver case, a head and neck case, and a prostate case with CTs and ROIs from the TROTS dataset. The data was imported into a research version of RayStation 2023B and proton arc plans were using a dynamic full revolution PBSArc with 180 energy layers were generated. Each case had three objective functions: a robust min function and robust max function on the PTV, and a non-robust dose fall-off function on the external. The thresholds for the min and max functions were set to the target dose $\pm 0.5$ Gy. The dose fall-off function was weighted 1000 times lower than the other functions. The robustness included 21 scenarios, created by 3 mm position uncertainty and 3\% density uncertainty. Each case was allowed to run for 100 iterations until spot filtering, and then for 50 iterations more. The target dose was 50 Gy for the liver case, 70 Gy for the prostate case, and 46 Gy for the head and neck case. Each case consisted of 30 fractions. The dose distribution of the prostate case can be seen in figure \ref{fig:prostate}.

The time required for an upward energy switch was assumed to be 5 s and 0.5 for a down switch. The spot switching time was set to 2 ms and the spot delivery is 5 ms for one MU. The jerk limited machine parameters were used, and the the gantry safety window was set to $1^\circ$.

Four velocity profiles were designed for each case. The first was produced by the ATOM algorithm, and the rest were produced by relaxing the first three assumptions and treating the problem like a continuous optimization problem with six variables per energy layer: the acceleration, velocity and position at the start and end of the irradiation of all layers. The velocity and acceleration before the first layer was set to zero, and the same thing was done to the velocity and acceleration after the last layer. The start and end position of the entire arc were set to 0 and 359. All variables in the optimization that correspond to positions had bounds between zero and one, where 1 corresponds to the smallest angle where an energy layer can be delivered and 1 corresponds to the largest angle. The velocity variables were bounded between 0 and 5, and the acceleration variables where bounded between 0 and 0.5.

The global optimization algorithm Dual Annealing \cite{genSimAnn} was used to create the second velocity profile. It used the SciPy \cite{SciPy} implementation and was allowed to run for 12 h per case. The third and fourth trajectories were created using a hill climbing optimization algorithm that randomly perturbed $0.5\%$ of the variables with a Gaussian perturbation with a standard deviation of 0.001. The difference between the third and fourth velocity profile was only the starting guesses: one used the solution found by ATOM as a starting guess and the other used a step and shoot velocity profile which came to a full stop for every energy layer. These algorithms were allowed to run for 12 h, too. For a fair comparison between Python scripts and C++ code, we compare the number of calls to the \textit{calculate} function in Ruckig rather than running time, since the calculate function is the computational bottle-neck.

\section{Results}
\subsection{Synthetic Data}
\begin{figure}[H]
    \centering
    \includegraphics[scale=0.5]{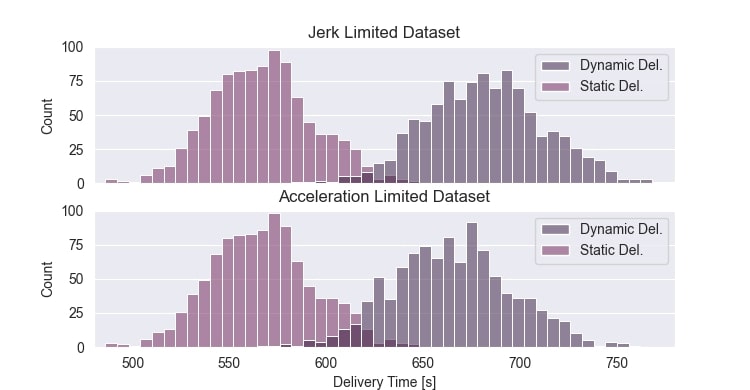}
    \caption{The dynamic and static delivery times of $2 \times 1000$ different random plans.}
    \label{fig:hist}
\end{figure}

For each plan we calculated the \textit{static delivery time} as the sum of the energy layer switches and the irradiation times, representing the beam delivery time if the gantry were fully static. This was subtracted from the total delivery time to give the \textit{dead time}. The static and dynamic delivery times, as calculated by ATOM, are shown in Figure \ref{fig:hist}. It should be noted that the dynamic delivery time is significantly slower than the static delivery one, which means that using the static delivery time as a model might be ill advised in some cases. The computational run times are given in Table \ref{tab:my_label2}. The algorithm is a lot faster with pruning than without in both datasets.

\begin{table}[H]
    \centering
    \begin{tabular}{c|c | c}
             & With pruning & Without pruning \\
            \hline
        Jerk Limited Datase & 0.70 s & 2.04 s\\
        Acceleration Limited Dataset & 0.46 s & 2.11 s

    \end{tabular}
    \caption{The mean computational run times for the plans in the two datasets. Each plan contains 360 energy layers.}
    \label{tab:my_label2}
\end{table}

\begin{figure}[H]
    \centering
    \includegraphics[scale=0.85]{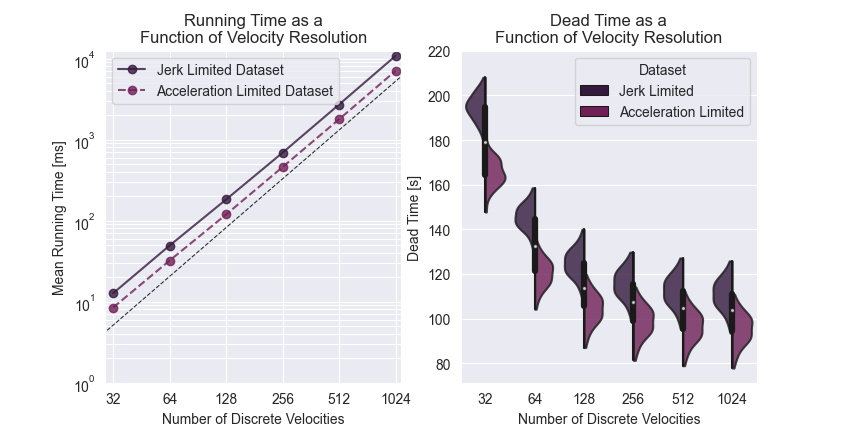}
    \caption{\textit{Left}: The running time as a function of the number of discrete velocities on a log-log plot. The thin bottom line is the function $x^2/c$, showing that the time complexity of ATOM is roughly quadratic with respect to the number of velocities. The constant $c$ was set to 200 to roughly mimic the scaling of the algorithm. \textit{Right}: The dead time is defined as the total delivery time minus the time of irradiation, spot switching and energy switching. This shows the estimated dead time, for all 2$\times$100 plans, using different velocity grids.}
    \label{fig:line}
\end{figure}

The results from the runs with varying velocity discretization are found in Figure \ref{fig:line}. The time complexity of ATOM with respect to the velocity resolution is roughly quadratic, which is to be expected given that the number of edges grows quadratically with the number of velocities (the number of edges is $O(NM^2)$ and the number of nodes is $O(NM)$). We think 256 different velocities is a decent trade-off between running time and accuracy, based on these results.

\begin{figure}[H]
    \centering
    \includegraphics[scale=0.5]{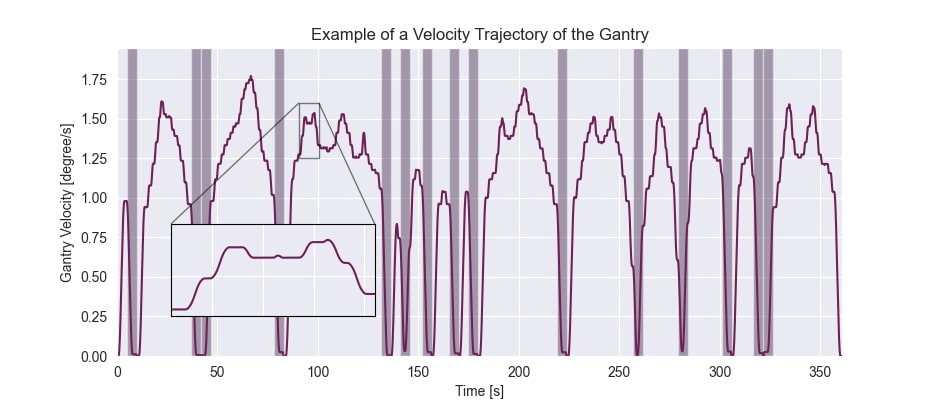}
    \caption{The velocity of the gantry for an example plan with 180 energy layers and 2 degrees between each energy layer. The dark areas symbolize the time required for upward energy switches. The machine parameters are $v_{max}=5$, $a_{max}=0.5$ and $j_{max}=0.5$. The zoomed-in window shows that the curve is smooth, and mostly consists of constant velocity.}
    \label{fig:exampleTraj}
\end{figure}

The gantry velocity curve of an example plan can be seen in Figure \ref{fig:exampleTraj}.

\subsection{Patient Data}
The global optimizer struggled to find a decent solution in all three cases, however it is likely that it would have done so given enough time. It did not find a feasible solution at all in the liver case. The local optimizer seems to have gotten stuck in a local minima, which was to be expected. However, when its initial solution was a start-and-stop arc it failed to find a solution with a delivery time faster than 900 s, for all three cases. This is roughly three times slower than the solution found by ATOM. The fact that both the hill climbing algorithm and the dual annealing algorithm failed to find a good solution in a reasonable time shows that designing these velocity profiles is a non-trivial problem and motives the use of the given assumptions.

\begin{figure}[H]
    \centering
    \includegraphics[scale=0.8]{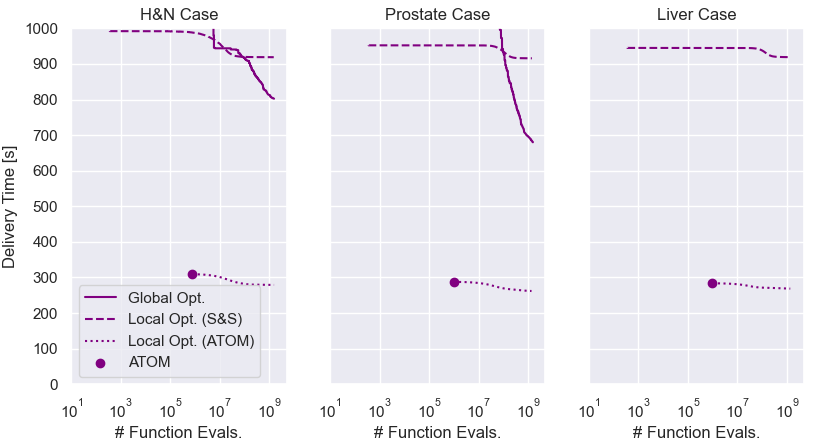}
    \caption{The delivery time as a function of number of Ruckig evaluations for thee patients and four methods. The ATOM case requires a certain number of Ruckig evaluations and is therefore represented as a single point. A local optimization algorithm was run with two different starting points: a velocity profile that starts and stops (S\&S) such that every energy layer is delivered with a static gantry, and the velocity profile generated by ATOM. Note that the global optimization was unable to find any feasible solutions for the liver case. All algorithms were given a maximum run time of 12 h.}
    \label{fig:human_result}
\end{figure}

The local optimization that ignored all our assumptions (except for the one about reversing gantry direction) and used the solution found by ATOM as its initial solution managed to improve on that solution in all three cases. In the head and neck case the delivery time was reduced from 309 s to 279 s. In the prostate case it went from 288 s to 266 s, and in the liver case it went from 284 s to 271 s.
\section{Discussion}
We have created a simple, fast and open-source algorithm that can estimate the fastest delivery time of an ion arc plan. The calculation time for a proton arc plan is under a second, using a resolution of 256 discrete velocities.

The solutions found by ATOM are optimal under the listed assumptions, but a local search around the solution found by ATOM showed that it was possible to reduce the delivery of 5 min plans by roughly 15-30 s.

Without our assumptions the naive solution would entail configuring each state in the graph to encompass the angular position, velocity, and acceleration of the gantry during the intervals between irradiation and energy switching. One could still assume that jerk is zero in all these instantaneous states while allowing non-zero jerk between them, since there are no bounds on the derivative of the jerk. This would result in six variables for each energy layer: position, velocity, and acceleration at the start and end of irradiation of an energy layer. With a discretization factor of $n=256$, this yields an approximate count of $256^{6} \approx 2.8 \cdot 10^{14}$ possible states per energy layer.

However, by introducing the assumption that the gantry won't accelerate during irradiation, we establish that the acceleration at the start and endpoints are zero, and the start and end velocities are identical. Furthermore, we can calculate the end position from the start position, velocity, and irradiation duration for the layer. Consequently, each layer requires only two variables: initial position and velocity. Incorporating the assumption that the gantry angle windows remains centered around the specified gantry angle further reduces the variables per energy layer to one: velocity. The necessary span of the layer's position is determined by multiplying velocity by the irradiation time, and by assuming symmetry in the position span around the specified gantry angle, the start and end positions of a layer can be calculated from its velocity. Thus, we only necessitate $n=256$ states per layer, rendering the problem computationally feasible.

However, this algorithm is just one part of the jigsaw puzzle. Using ATOM to compute delivery time for a real plan requires the user to provide a delivery model of the relevant machine, providing accurate irradiation time per layer as well as energy layer switching times between layers. However, these kinds of models already exist for IMPT machines and are starting to be developed for PAT machines \cite{model, pfeiler}.

It is worth noting that a faster algorithm would open up the possibility to take dynamic beam delivery time into account during the creation of a treatment plan. For example, algorithms such as ELSA \cite{elsa} and the MCO-based algorithm by Wuyckens et al. \cite{pareto} design plans with short delivery times by minimizing the number of upward energy switches. Further reduction in the delivery time may be achievable by tailoring the time estimate to a specific treatment machine and optimizing it directly accounting for the specific characteristics and capabilities of the treatment machine in use. However, this would require thousands, if not tens of thousand, of evaluations of ATOM which would unfortunately give a run time on the order of an hour. Another possibility would be to add a penalty/constraint on the BDT during the spot weight optimization, but this could also significantly increase the running time of the optimization. We therefore leave the following open research question: \textit{Is it possible to significantly decrease the computational time of ATOM without sacrificing the fine velocity discretization?}

A second open research question concerns the stability of the velocity profile generated by ATOM. For real plans one never gets exactly the expected delivery time per layer, or the expected velocity at a point. ATOM does not consider these uncertainties and might be overly sensitive to them. The second open question is thus: \textit{How sensitive is ATOM generated trajectories to small perturbations of irradiation time, layer switching time, and velocity? And how could one adjust ATOM to handle these uncertainties?}

\section{Conclusion}
We have developed ATOM, an open-source algorithm for finding velocity profiles for delivery of ion therapy treatment. ATOM computes the gantry path in under a second and considers constraints on the magnitude of velocity, acceleration, and jerk. The gantry velocity profile generated allows for fast delivery of ion arc plans -- in fact, the delivery time is minimized to optimality with respect to our modelling assumptions. Furthermore, ATOM enables quick evaluation of delivery times for a variety of candidate plans, giving planners the option to take delivery time into consideration.

\section{Acknowledgements}
We thank Ivar Bengtsson for discussions on optimization methods, and Sophie Wuyckens and Erik Engwall for valuable comments on the manuscript.




\addcontentsline{toc}{section}{\numberline{}References}








\bibliographystyle{unsrt}  
\bibliography{templateArxiv}

\end{document}